\documentclass[preprint]{aastex}
\usepackage{rotating}
\usepackage{amsmath}		
\usepackage{timing}
\usepackage{ISM}
\usepackage{natbib}
\usepackage[margin=26mm]{geometry}
\newcommand{\tp}{t^{\prime}}

\newcommand{\NX}{N_X}
\newcommand{\Np}{N_p}
\newcommand{\Ns}{N_s}
\newcommand{\varrat}{\psi}

\newcommand{\Chat}{\widehat C}

\newcommand{\CF}{CCF}
\newcommand{\CFs}{CCFs}

\newcommand{\siggw}{\sigma_{\rm gw}}
\newcommand{\sigrn}{\sigma_{\rm r}}
\newcommand{\sigwn}{\sigma_{\rm n}}
\newcommand{\snrccf}{S}
\newcommand{\zetaM}{\xi_M}	
\newcommand{\monemin}{{m_1}_{\rm min}}
\newcommand{\mtwomax}{{m_2}_{\rm max}}

\begin{document}
\singlespace

\parindent 0pt
\newcommand{\be}{\begin{eqnarray}}
\newcommand{\ee}{\end{eqnarray}}
\newcommand{\etal}{et al.}

\title{Minimum Requirements for Detecting a Stochastic Gravitational Wave
	Background Using Pulsars}
\author{J. M. Cordes }
\affil{Astronomy Department, Cornell University} 
\email{cordes@astro.cornell.edu}
\and
\author{R. M. Shannon}
\affil{Astronomy Department, Cornell University; CSIRO Astronomy
and Space Science, Epping, NSW, 1710, Australia } 
\email{ryans@astro.cornell.edu}

\begin{abstract}
We assess the detectability of a nanohertz gravitational wave (GW) background 
with respect to
additive white noise and especially red noise in the timing of 
millisecond pulsars.
We develop detection criteria based on the shape and amplitude 
of the cross-correlation function summed over pulsar pairs in 
a pulsar timing array. 
The distribution of correlation amplitudes
is found to be non-Gaussian and highly skewed, which  significantly influences
the detection and false-alarm probabilities.
When only white noise combines with GWs in timing data, our detection results
are consistent with those found by others.  Red noise, however, drastically
alters the results. 
We discuss methods to meet the challenge of GW detection 
(``climbing mount significance'') by distinguishing between GW-dominated
and red or white-noise limited regimes. 
We characterize plausible detection regimes by evaluating 
the number of millisecond pulsars that must be monitored in a high-cadence,
5-year timing program for a GW background spectrum
$h_c(f) = A f^{-2/3}$ with $A = 10^{-15}$~yr$^{-2/3}$.
Our results suggest that unless a sample of
20 super-stable millisecond pulsars can be found --- 
those with timing residuals from red-noise contributions 
$\sigma_r \lesssim 20$~ns ---  a much larger timing program
on $\gtrsim 50 - 100$ MSPs will be needed.
For other values of $A$, the constraint is 
$\sigma_r \lesssim 20~{\rm ns}~(A/10^{-15}~{\rm yr}^{-2/3})$.
Identification of suitable MSPs
itself requires an aggressive survey campaign followed by characterization
of the level of spin noise in the timing residuals of each object. 
The search and timing programs 
will likely require substantial
fractions of time on new array telescopes in the southern hemisphere
as well as on existing ones.
\end{abstract}

\section{Introduction}
There is current strong interest in exploiting the spin
stability of millisecond pulsars (MSPs) to detect
gravitational waves (GWs) at nanohertz frequencies ($\sim 0.1-1$~cy~yr$^{-1}$).
Sources of GWs in this band include mergers of supermassive black holes that collectively produce
an isotropic background 
\citep[][]{2003ApJ...583..616J, 2004APS..APR.L9002P, 2005ApJ...625L.123J, 2008MNRAS.390..192S} 
and in a few cases may be detected individually 
\citep[][]{2001ApJ...562..297L, 2004ApJ...606..799J, 2010ApJ...718.1400F, 2010MNRAS.407..669Y}.
Timing may also detect GW backgrounds from cosmic strings
\citep[][]{2010PhRvD..81j4028O}, 
the influence of massive gravitons
\citep[][]{2010ApJ...722.1589L}, 
or solar system perturbations from primordial black holes
\citep[][]{2007ApJ...659L..33S}.
Detection methods have been based on finding excess variance 
in the timing residuals of individual sources, 
investigating spectral signatures in power spectra,
or identifying the angular correlation expected 
between pulsar pairs from GWs passing through the solar system. 
Astrophysical and instrumental processes
limit the timing precision of any given pulsar and the overall
sensitivity of a pulsar timing array (PTA).

However, the efficacy of detection
methods has received very uneven assessment with respect to contamination
from different kinds of  additive noise.  These include both  
white and red noise processes, the latter having 
power strongly  concentrated at lower fluctuation frequencies.    
We evaluate their impact on the sensitivity to a 
stochastic GW  background, which itself comprises a red noise process.    

In this paper, we are concerned with {\em detection} of GWs as distinguished
from their detailed characterization.
Recent work has considered both frequentist 
\citep[e.g.][]{2005ApJ...625L.123J, 2011arXiv1102.2230Y}
and Bayesian approaches 
\citep[e.g.][]{2009MNRAS.395.1005V, 2011arXiv1103.0576V}
to the detection problem. 
While these methods are
robust to varying degrees,
our view is that detection needs to be corroborated with convincing diagnostics.
We draw an analogy with the detection of a new spectral line
or detection of cosmic microwave background fluctuations.   
Bayesian inference can yield the best probabilistic constraints on 
signal parameters, 
but most observers are convinced
of an underlying detection and characterization only with the display of
a spectral line or the power vs. 
spherical harmonic number that indicates a significant signal with respect
to measurement errors.       
The corresponding quantity  in the pulsar-GW problem 
is the cross correlation between the timing residuals of pulsar pairs
or some related quantity.

We derive a general expression for the signal-to-noise ratio (SNR)
of a correlation-based detection statistic  and develop a 
detection protocol based on the shape and amplitude of the  
cross-correlation function. 
We assess the challenges for a likely detection and estimate
the minimum number of MSPs needed under different circumstances. 
To do so, 
we consider a hypothetical pulsar distribution that yields the
highest possible SNR for the correlation function, 
all else being equal. This is a configuration where 
$\Np$ pulsars are in the same direction but at different distances
so that the perturbation from  GWs passing through
the solar system is 100\% correlated between all objects. 
Any other configuration will yield smaller correlation and SNR.

The correlated effect on times of arrival (TOAs) for different pulsars is
produced by GWs passing through the solar system
\cite[][]{1983ApJ...265L..39H}. 
We designate the Hellings and Downs angular correlation function as 
$\zeta(\theta)$ for two objects separated
by an angle $\theta$, with normalization $\zeta(0) = 1$;    
this differs from other definitions in the literature that include
a delta function associated with GWs at the pulsar location, which
yields $\zeta(0^+) = 1/2$. For our purposes, we keep the pulsar term separate.

We assess detectability in terms of different levels of white and red noise
in timing residuals.   Our work follows
\citet[][]{2010arXiv1010.3785C} where we assess a wide range
of  contributions to TOA errors from the pulsar, the interstellar medium, 
and from instrumental effects.  White noise not only includes radiometer
noise but also pulse-to-pulse phase jitter from magnetospheric activity
and from an effect associated with interstellar scintillation, which are
pulsar and line-of-sight (LOS) dependent, respectively.     
We have also shown that red spin noise, which is common in 
canonical pulsars --- those with periods of order one second and surface 
magnetic fields $\sim 10^{12}$~G --- is also  
to be expected in MSPs but at low levels in accord with their spin 
parameters \cite[][]{2010ApJ...725.1607S}.     
Interstellar scintillation also contributes red-noise TOA perturbations,
some of which can be corrected before any analysis for GWs. 

In the next section we describe the cross correlation analysis
of a simplified timing model and develop detection criteria for assessing the
presence of a GW signal.  
In \S~\ref{sec:mount} we describe how the prospects for detection can be maximized.
We summarize our results and discuss them in broader terms in \S~\ref{sec:discussion}.
The Appendix defines quantities used to characterize the timing residuals  and describes
simulations.

\section{Cross Correlation Detection}
\label{sec:detection}

We use a simplified model for the timing residuals of a pulsar 
by excluding real-world effects such as time transfer and the error in the 
location of the solar-system barycenter 
\citep[e.g.][]{1986ARA&A..24..537B}.
\be
x(t) = e(t) + u(t), 
\label{eq:xmod}
\ee
where $e(t)$ is the ``Earth'' part of the gravitational wave (GW)
background and $u(t)$ includes all other processes, which we assume
are uncorrelated between different pulsars.   At minimum, $u(t)$ includes
the GW perturbation $p(t)$ at the pulsar's location that acted on the
measured signal a time $D/c$ earlier, where $D$ is the pulsar's distance.
Later we expand $u(t)$ into three components that are uncorrelated
with $e(t)$ and with each other,
\be
u(t) = p(t) + r(t) + n(t),
\label{eq:xmod2}
\ee
where $r(t)$ is red noise associated with spin noise (``timing noise'')
in the pulsar or with multipath propagation in the interstellar
medium (ISM), and $n(t)$ is white noise that represents measurement 
errors of different kinds.   Each is characterized by a correlation function
$\sigma_x^2(T) \rho_x(t,t^{\prime})$, where $\sigma_x^2(T)$ is the ensemble-mean
variance over an interval $[0,T]$ and $\rho_x(t, t^{\prime})$ is the normalized
correlation function defined with two arguments to handle non-stationary
as well as stationary processes. 

\begin{figure}[h!]
\begin{center}
\includegraphics[scale=0.40, angle=0] 
{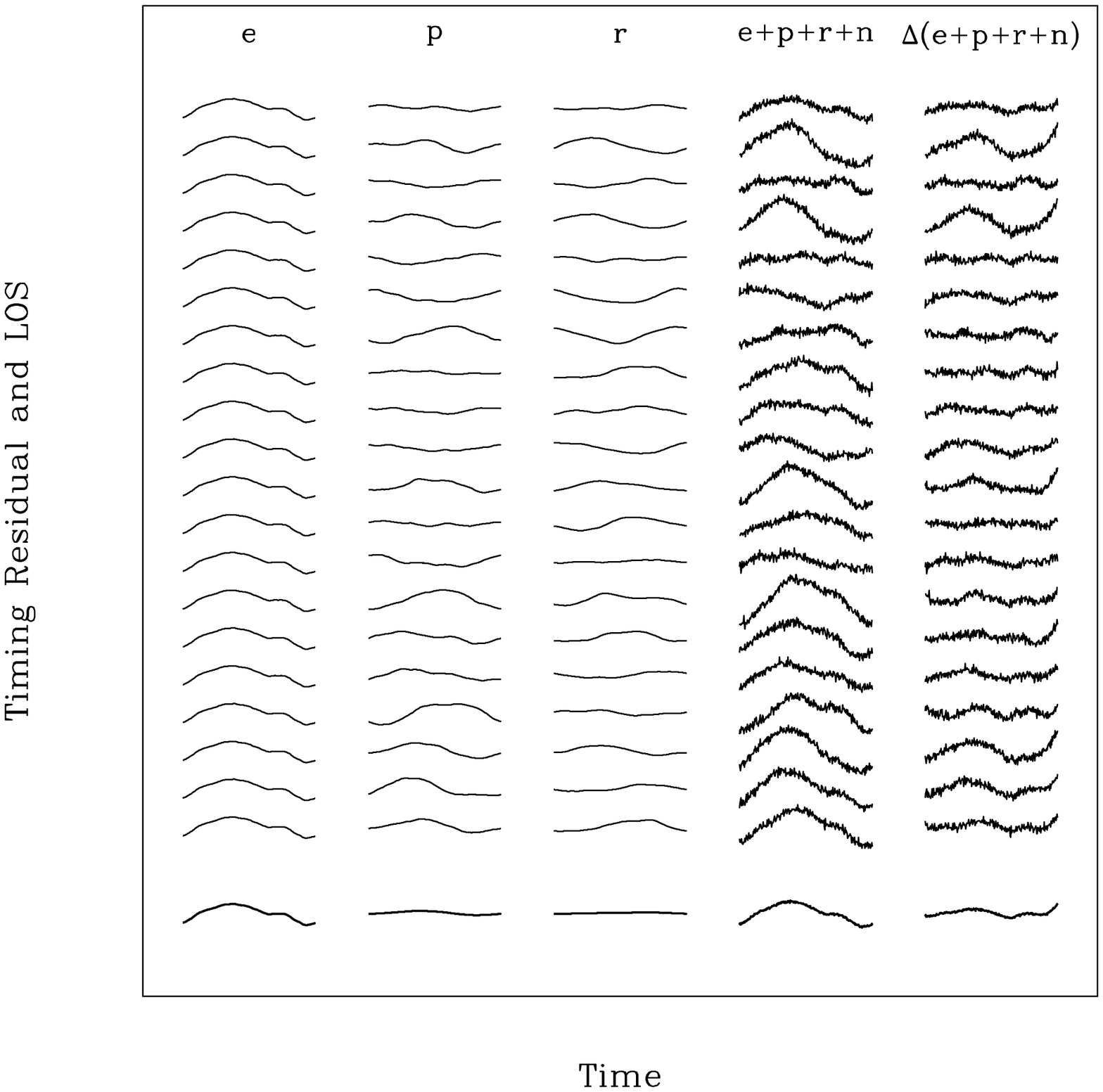}
\includegraphics[scale=0.40, angle=0] 
{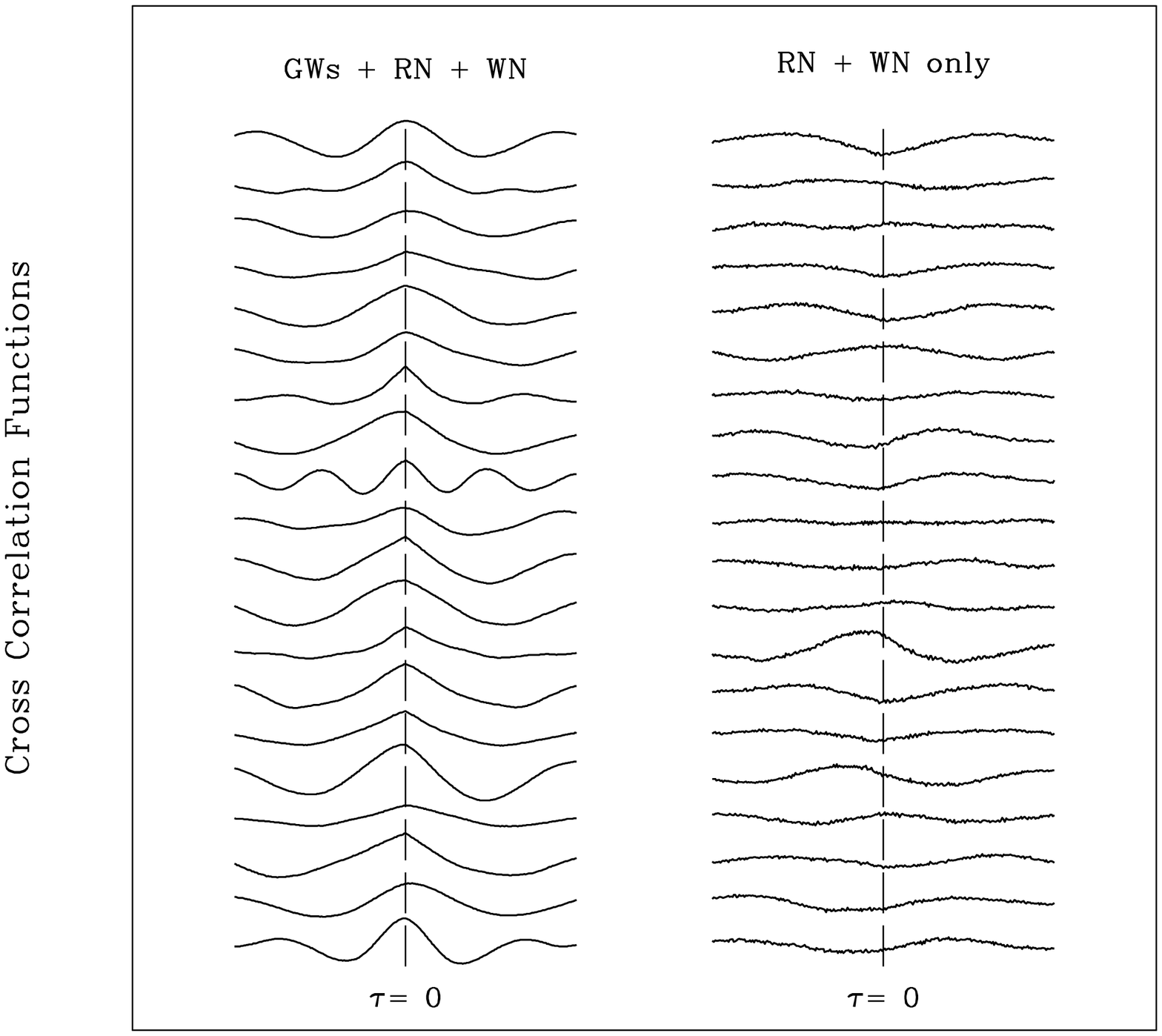}
\caption{
(Left) Simulated time series for a 20-pulsar timing array over 5 years. 
The columns from left to right show
the $e, p$ and $r$ terms in the timing model, their sum added to
white noise ($n$), and the residuals $\Delta(e+p+r+n)$ from a second-order polynomial fit.
The simulations include
GWs, red noise and white noise that have identical rms values 
$\siggw = \sigrn = \sigwn = 20$~ns.  
in the post-fit time series. 
The bottom row shows the
average of each column over the 20 pulsars.  
(Right) \CFs\ for a 20-pulsar PTA after removing a quadratic fit to each
residual time series before cross correlating between all pairs.    
The left-hand column shows results for 
20 realizations of the PTA shown in the left-hand panel.  
The right-hand column shows cases with no GW contribution.
\label{fig:ccfs}
}
\end{center}
\end{figure}

\begin{figure}[h!]
\begin{center}
\includegraphics[scale=0.40, angle=0] 
{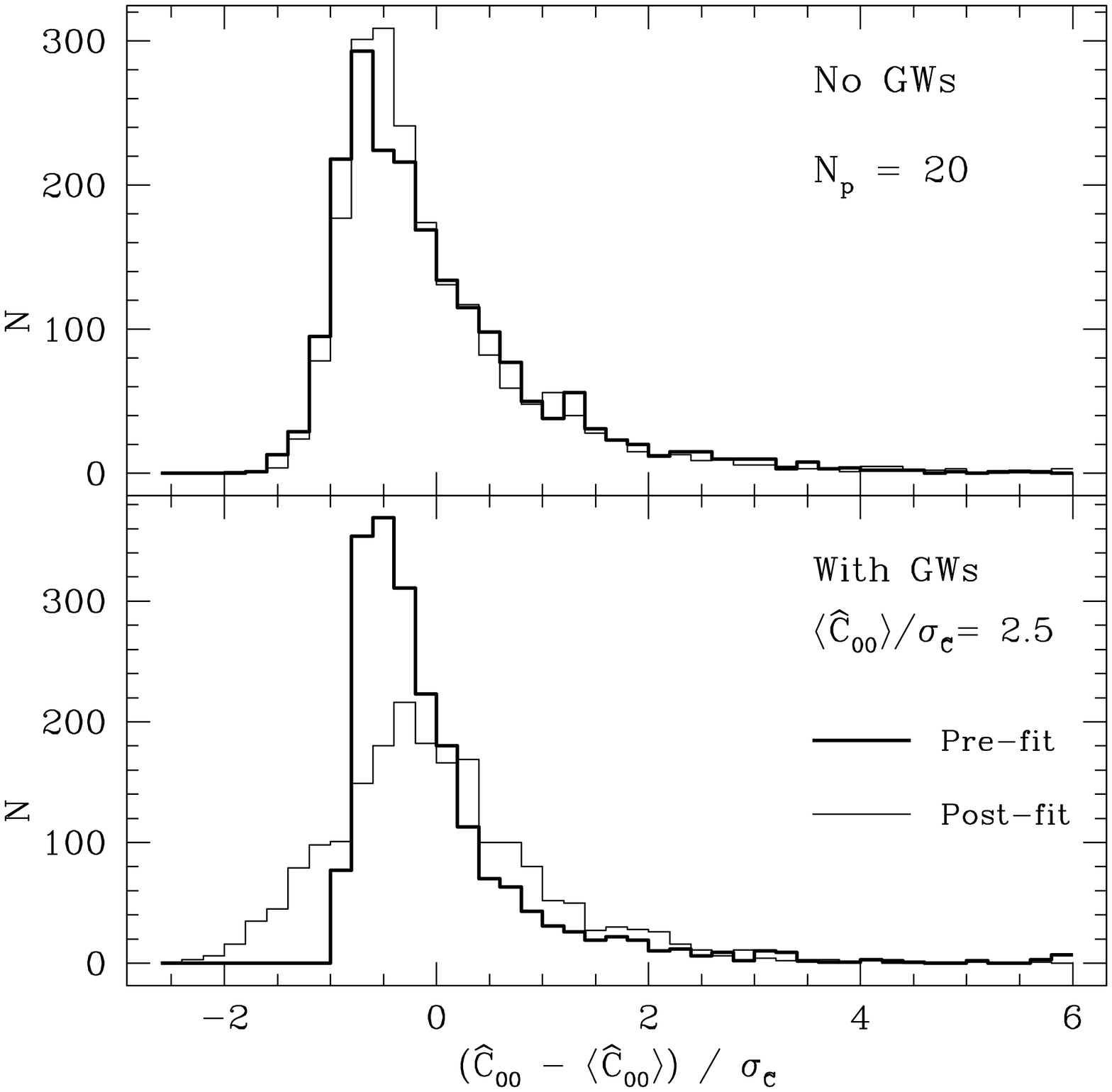}
\includegraphics[scale=0.40, angle=0] 
{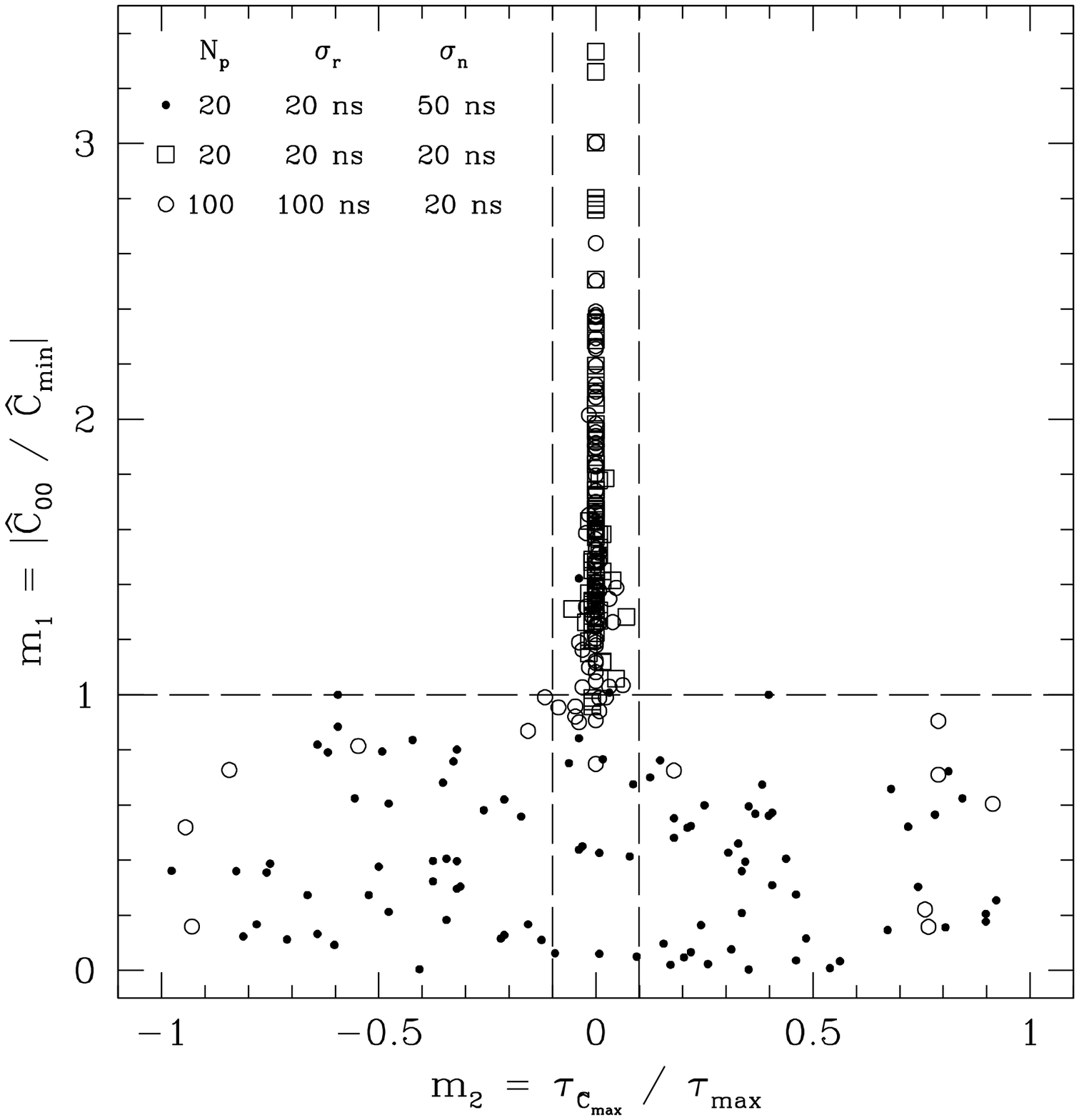}
\caption{
(Left)
Histograms of the zero-lag values of cross-correlation functions based
on simulations of PTAs that use 20 pulsars. The mean of 2000 realizations is
subtracted from each value and the result is normalized by the 
rms deviation, $\sigma_{\Chat}$.  
Heavy lines show the histogram of $\Chat_{00}$ calculated using the
pre-fit time series and the light lines  show post-fit results.
In the top panel GWs do not contribute to the timing residuals 
while in the bottom panel they contribute
$\siggw = 20$~ns after a polynomial fit.
Uncorrelated red and white noise contribute
post-fit rms values $\sigma_r = \sigma_n = 20$~ns  to each time series.
Note that the the histograms in the bottom panel have been shifted
by the average, post-fit  signal to noise ratio, which for this case is
$S = \langle \Chat_{00} \rangle / \sigma_{\Chat} = 2.5$,
where $\sigma_{\Chat}$ is calculated from the simulations.
The histograms in the upper panel have $\langle \Chat_{00} \rangle = 0$.
\label{fig:c00hists}
(Right)
Plot of metrics $m_1$ vs. $m_2$ that characterize the shape of the cross 
correlation function of the post-fit timing residuals, as defined in the text;  
points are shown for 100 realizations of each case. 
The dashed lines denote the minimum threshold for the vertical axis and the
maximum departure of the \CF\ peak from zero lag in order to provide a plausible
detection.    
The open circles are for a case with no GW contribution while the open squares
and filled circles have a post-fit GW rms of 20~ns in each time series.
\label{fig:m1vsm4}
}
\end{center}
\end{figure}

Figure~\ref{fig:ccfs} (left panel) shows simulated time series for a 20-MSP PTA.
The time series show the $e$ terms to be identical
for all 20 pulsars, as assumed, while the $p$ and $r$ terms are different.  
The sum of all terms and the residuals from a second-order polynomial fit are
also accordingly different.   
The bottom row of the panel shows sums of the various terms in the 
first five columns, which demonstrate the reduction in rms by $1/\sqrt{20}$
for the $p$ and $r$ terms.   
In the right-hand panel
\CFs\  for 20 realizations of the PTA are shown where 
the GW, red-noise and white-noise
contributions are equal over the data span $T$,   
corresponding to $\psi = 1$ and
$\sigma_e = \sigma_p = \sigma_r = \sigma_n$ in one set of curves (the left
column), while the GWs are turned off for the curves in the right-hand column.

The model in Eq.~(\ref{eq:xmod})-(\ref{eq:xmod2}) is a sum
of Gaussian distributed  processes because all terms originate from conditions
that usually will  satisfy the central limit theorem (CLT).  
Focusing events in the interstellar medium 
\cite[][]{2010ApJ...717.1206C}
or events intrinsic to the pulsar can plausibly induce non-Gaussian statistics 
in data from some objects. 
The cross-correlation between pulsars, being a second moment, is a sufficient statistic for the 
angular correlation function.
We define the general temporal cross-correlation function (\CF) in the Appendix
and focus attention here on its zero-lag value
\be
\Chat_{00} \equiv \Chat(\theta = 0, \tau = 0) 
	=  
	\frac{1}{\NX }
	\sum_{i<j} 
	\hat C_{ij}(\theta_{ij}=0, \tau=0)
	= 
	\frac{1}{\NX T}
	\sum_{i<j} 
	\int_0^Tdt\, x_i(t) x_j(t),
\label{eq:Chat}
\ee
which defines the single-pair \CF\ $\hat C_{ij}$ and 
the number of unique pairs in the double sum over $i,j$  is
$\NX=\Np(\Np-1)/2$ for $\Np$ pulsars. 
We have used continuous time notation; we justify this in the Appendix where we also discuss
how to treat discretely sampled data.
The ensemble average is
\be
\left\langle \Chat_{00} \right\rangle  
	= \left\langle T^{-1} \int_0^T dt\, e^2(t) \right\rangle
	\equiv \varrat \siggw^2(T).
\label{eq:Chatave}
\ee
We use $\varrat$ to denote the ratio of the variances of the
actual Earth term  and the ensemble average variance.   For red 
processes with power-law spectra $\propto f^{-y}$, 
the range of $\varrat$ covers one to
a few {orders of magnitude}, with steeper power laws showing a
wider range, as discussed in the Appendix.   
This implies that 
the GW background yields an  actual rms 
TOA perturbation that can vary by more than a factor of ten. 
We define the signal to noise ratio 
$S=  \Chat_{00} / \sigma_{\Chat}$ in \S~\ref{sec:ccfsnr}.
Our definition for $S$ has some similarity to that defined by
\citet[][]{2005ApJ...625L.123J} but with a crucial difference.  
Their Eq.~4 is essentially a matched filter based on the 
Hellings and Downs angular correlation. 
In our notation, the numerator of their equation is
(with subscript ``J'' for Jenet~et~al.)
\be
\hat C_J = 
	\frac{1}{\NX }
	\sum_{i<j} 
	\left [\hat C_{ij}(\theta_{ij}) - \overline{C}\right]
	\left [\zeta_{ij} - \overline{\zeta} \right].
\ee 
The angular separation of the $i$-th and $j$-th objects is
$\theta_{ij}$ and 
barred quantities are sample means over all pulsar
pairs of $\hat C_{ij}$ and $\zeta_{ij} \equiv \zeta(\theta_{ij})$,
respectively.  The ensemble mean is 
\be
\langle \hat C_J \rangle
	= 
	\siggw^2 \left(\overline{\zeta^2} - {\overline{\zeta}}^2 \right),
\ee 
where
\be
\overline{\zeta^n} = \frac{1}{N_X} \sum_{i<j} \zeta^n(\theta_{ij}).
\ee
For the compact pulsar configuration we consider, $\langle \hat C_J\rangle$
vanishes because $\overline{\zeta^2} = \overline{\zeta} = 1$ 
and thus cannot be used to quantify detection in this case.  
If we redefine the weighted correlation as 
\be
\hat C_J^{\prime} &=& 
	\frac{1}{\NX }
	\sum_{i<j} 
	\hat C_{ij}(\theta_{ij}) \zeta_{ij}, 
\ee
the
Jenet et~al. test statistic becomes
\be
S_J^{\prime} &=& \frac{\hat C_J^{\prime}}{\sigma_{C_J^{\prime}}},
\ee 
which is identical to our definition for $S$
when the pulsar configuration is compact with  $\theta_{ij} = 0$. 

Later in the paper we will relate our results to an arbitrary configuration
of pulsars by considering $S_J^{\prime}$, which simply multiplies our
result for $S$ by the mean-square angular correlation over the sample,
$\overline{\zeta^2}$.

\subsection{Detection Criteria}
\label{sec:criteria}

Over an ensemble, the \CF\ vanishes unless there is a significant
correlated term from the $e(t)$ term 
(or from errors in the location of the solar system barycenter or from instrumentation 
that we do not include in our analysis). 
However, deviations from ensemble-average statistics in real data
will produce both false positive and false negative detections from
the uncorrelated terms in the timing residuals, 
like those shown in Figure~\ref{fig:ccfs}.

A detection protocol for GWs can exploit the following aspects of
timing residuals and their correlations: 
\begin{enumerate}
\item The timing residuals must include a red-noise process caused by 
one or more of the predicted isotropic GW backgrounds 
\citep[][]{2005ApJ...625L.123J}.
If GWs from any discrete source are significant,
there should be a corresponding departure from white-noise statistics 
described by a 
spectrum that depends on the nature of the GWs 
\cite[][]{2001ApJ...562..297L}.
\item The maximum of the \CF\ is at or near zero time lag, 
$\tau\approx 0$, depending on how strong the GWs are relative to
other contributions.
Uncorrelated contributions produce estimation errors in $\Chat_{00}$ 
that peak at arbitrary time lags in estimates using a finite number of pulsars 
and thus can induce false non-detections and false positive detections.
\item The zero-lag amplitude of the \CF\ must 
be significantly larger than expected when only uncorrelated terms contribute
to the time series. 
\item The correlation between pulsars is consistent with 
that expected (e.g. Eq. 5 of \nocite{1983ApJ...265L..39H} Hellings \& Downs,
1983)  
for an isotropic background or 
the equivalent angular correlation for
a discrete source. 
\item Any correlation established using one set of pulsars can be checked using a completely
independent set of pulsars.
\end{enumerate}

When white and red-noise processes are significant, the
estimated \CF\ has a high probability of peaking
at a non-zero time lag. If white noise dominates
the timing residuals of all pulsars, the \CF\ itself will vary rapidly with
time lag  and its formal maximum could be at any lag.   Red noise
by definition has a long correlation time so the 
\CF\ can appear quite smooth and yet peak far away from the origin. 
After a second order polynomial fit, the red-noise residuals will typically 
have two zero crossings so the \CF\ maximum is likely to be more centered than
the pre-fit \CF, an effect we see in simulations.  
Nonetheless, the full CCF provides important statistical tests of the zero lag value.



Figure~\ref{fig:c00hists} (left-hand panel) shows histograms of 
$\Chat_{00}$ obtained from simulations 
with and without a GW contribution and for both pre-and-post-fit cases.  
Under relevant circumstances, 
the distribution is asymmetric, with a long tail  for  
positive values while the mode and median are less than the mean.  
This counterintuitive result, discussed in the Appendix, occurs when the
correlated quantity includes a red process with a steep power spectrum. 
The time series for a red process is effectively dominated by
approximately one independent fluctuation so that the calculation 
of the \CF\ for one pair of pulsars does not satisfy the requirements
for the CLT to apply, as it would if the time series were only white
noise.  The sum over all pulsar pairs also does not satisfy the requirements 
in part because the number of pairs exceeds the number of individual objects.
Simulations show that 
the distribution for $\Chat_{00}$ becomes increasingly skewed for
steeper spectra and for larger $\Np$.  We have not yet explored if there is
an asymptotic form for the distribution.  The skewness we have identified
is similar in cause to that identified for studies of non-Gaussianity of
temperature fluctuations in the cosmic microwave background    
\citep[][]{2011arXiv1104.0930S}.

The long tail influences the false-alarm rate sigificantly.  One way to get
a more symmetric distribution is to average multiple estimates of the cross
correlation function.    Multiple estimates can be obtained by subdividing 
timing residuals into $M$ blocks, each  of length $T/M$, as mentioned
in \cite[][]{2010ApJ...725.1607S}. The CLT will apply to the average
so the distribution should tend to a Gaussian form.   We discuss this 
further in \S~\ref{sec:ccfsnr}.

\subsection{\CF\ Based Detection} 
\label{sec:cfdetection}

We define two metrics that characterize the shape of  the \CF, 
\be 
m_1 = \left \vert \frac{\Chat_{00}} {\Chat_{\rm min}} \right\vert,
\quad\quad\quad\quad
m_2 = \frac{\tau_{\Chat_{\rm max}}} {\tau_{\rm max}};
\ee
$m_1$ is the ratio of the \CF\ at zero lag to the most negative value;
$m_2$ is the offset of the \CF\ maximum relative to 
the maximum calculated lag, which in our simulation is 
$\tau_{\rm max} = T/2$.  The two metrics characterize the shapes of the
\CFs\ without reference to the actual signal to noise ratio of the GW signal
and therefore provide a direct empirical mechanism for assessing 
the presence of a GW signal.
We later define the
signal to noise ratio $S$ as an ensemble average quantity 
that can be related to physical models for the GWs.   
The metric $m_1$ is a similar
measure but is based on a single realization of the correlation estimate and is normalized
by the minimum  of the particular \CF, not the rms value over an ensemble. 
Figure~\ref{fig:m1vsm4} (right panel) 
shows  a scatter plot of $m_1$ and $m_2$ from
simulations that displays a peak in $m_1$ near $\tau = 0$.   The
fraction of points in the peak depends on the strength of the
GWs compared to other contributions and on the number of pulsars.   
For residuals dominated by uncorrelated red noise, there is a sizable fraction
($\sim 15$~\%) of cases with peak values that can mimic a GW
detection.
We define a joint detection criterion that comprises
a lower bound  $m_1 > \monemin$ and an upper bound 
$\vert m_2 \vert \le \mtwomax$. 
We also define the detection fraction 
$f_d$ as the fraction of PTA realizations
in a simulation that satisfy the detection criteria.   
The corresponding false-alarm fraction is defined as the detection fraction 
in the absence of any GW signal. 

\begin{figure}[h!]
\begin{center}
\includegraphics[scale=0.40, angle=0] 
{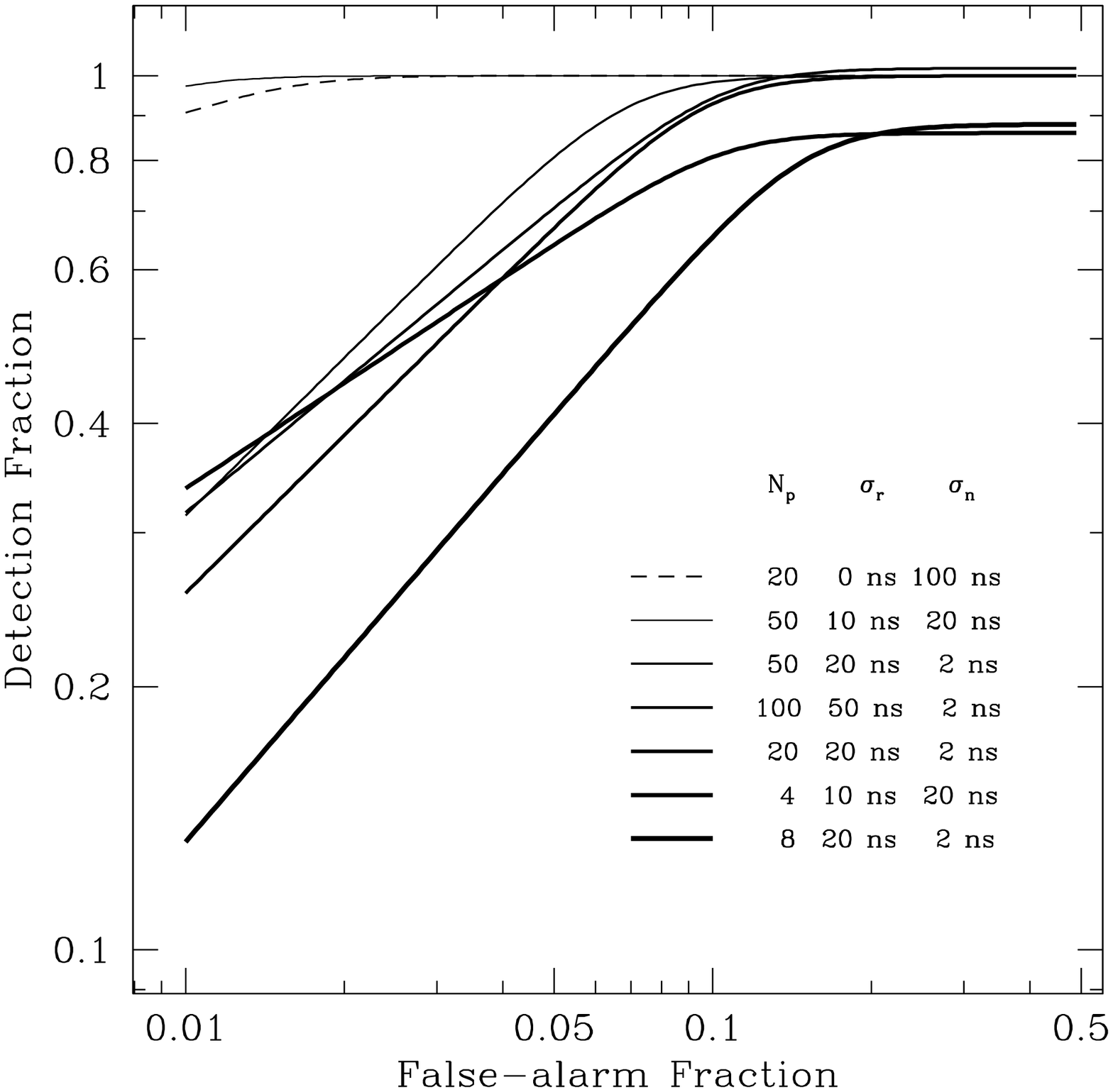}
\includegraphics[scale=0.40, angle=0] 
{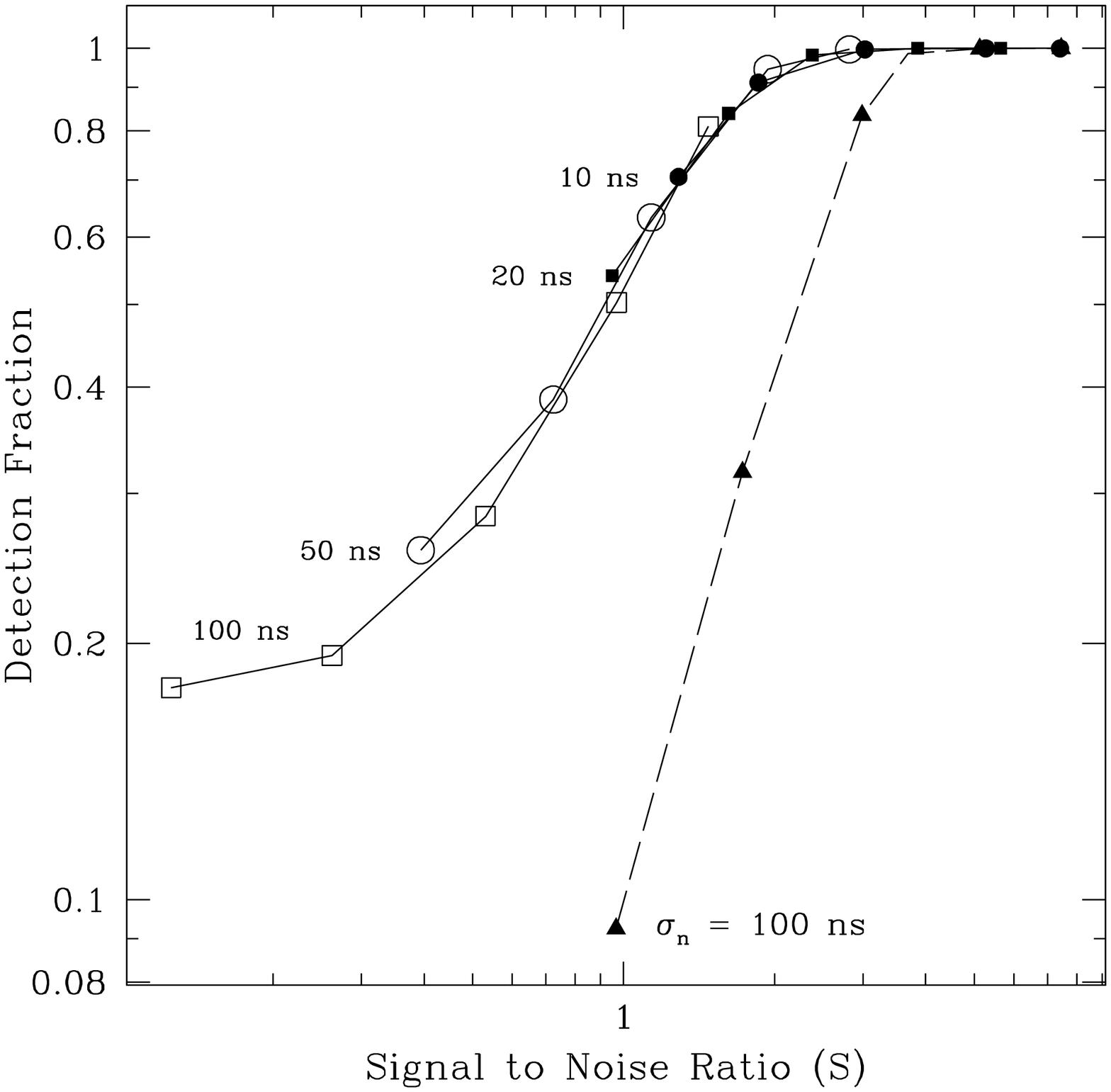}
\caption{
(Left)
ROC  curves showing detection fraction vs. false-alarm fraction for
different detection criteria   obtained by varying
the thresholds for $m_1$ and $m_2$.   
The detection fractions were calculated 
for the same GW strength (20~ns over 5~yr) but different numbers of
pulsars and for different levels of
red and white noise noise,  as labeled.  
The false-alarm fraction 
was obtained by turning off the GW signal 
and keeping the red and white-noise levels the same as for the
GW ``on'' case.
\label{fig:rocproc}
(Right) 
Detection fraction plotted against the expected signal-to-noise ratio, $S$, of
the cross correlation function.  The five points for each curve correspond to
PTAs with $\Np = 4,8,20,50$ and 100 pulsars.   
Solid lines: The rms red and white noise
are equal as labelled for each curve near the point corresponding to $\Np=4$.  
Dashed line: A case with white noise with 100~ns rms  (no red noise) 
added to the GW signal.  
\label{fig:fdvsSNR}
}
\end{center}
\end{figure}

Inspection of Figure~\ref{fig:m1vsm4}  suggests 
a threshold $\monemin = 1$, which  
is plausible since the \CF\ of a noise-only signal is likely to have approximately
equal positive and negative-going excursions.   
Also consistent with the figure is a threshold  $\mtwomax = 0.1$,
which enforces the zero time-lag nature of the ``Earth'' part
of the  GW signal but rejects cases where noise processes steer the 
\CF\ maximum away from zero lag.  

The right-hand panel in Figure~\ref{fig:ccfs} shows cases where the
\CFs\ satisfy the detection criteria (large amplitude and peak at $\tau = 0$)
and others that do not.  Conversely, when there is no GW contribution,
red noise can cause false positive cases that satisfy the criteria.  
We quantify these features in the discussion that follows.

Specifying a detection criterion requires consideration of the
tradeoff between
the detection fraction, $f_d$, and the false-alarm fraction, $f_{fa}$.
Figure~\ref{fig:rocproc} (left panel) shows ``ROC'' (receiver operating characteristics) curves calculated by varying
$\monemin$ and $\mtwomax$ to alter the detection and false-alarm fractions.  
Each curve corresponds to
a particular PTA (number of pulsars and levels of red and white noise). 
Ideally, one would like to have 100\% detection fraction with no false alarms.
The cases 
$(\Np, \sigrn, \sigwn) = (50, 10~{\rm ns}, 20~{\rm ns})$ 
and
$(\Np, \sigrn, \sigwn) = (20, 0~{\rm ns}, 100~{\rm ns})$ 
come closest to this ideal. 
All of the other cases, which have larger noise levels
or  smaller numbers of pulsars in the PTA depart significantly from the ideal.
With 20 pulsars having 20~ns of red noise and negligible white noise, 
a 90\% detection fraction comes at the expense of a 11\% false-alarm fraction.
A larger number of pulsars (such as 50 pulsars with 20~ns red noise or 
100 pulsars with 50~ns of red noise) decreases the false-alarm 
fraction to 8\%.    


The mapping of signal-to-noise ratio $S$ and detection fraction is shown
in Figure~\ref{fig:fdvsSNR} (right panel). 
Most of the curves shown are for equal levels
of red and white noise and different pulsar numbers (solid curves).   
Detection fractions $f_d \gtrsim 0.8$ require $S > 1.5$
and $f_d \gtrsim 0.95$ requires $S\gtrsim 2$.  
A case with 100~ns white noise (dashed curve) shows that $\Np = 20$ pulsars
yields $S\approx 3$ and a detection fraction $>80$\% and minimal
false-alarm fraction as shown in the left-hand panel.   Our results are
therefore broadly consistent with those of \citet[][]{2005ApJ...625L.123J}, who
consider only white noise TOA errors.   The primary conclusion from
Figure~\ref{fig:fdvsSNR} is that red noise drastically alters the detection
and false-alarm fractions and therefore also any assessment of GW
detectability.

\subsection{SNR of the Zero-lag Cross Correlation}
\label{sec:ccfsnr}

The detection criteria defined above are based solely on the shape
of the cross-correlation function. It is useful to relate the detection fraction to the
ensemble-average SNR of the correlation function (at zero lag), because the SNR
can be related to the GW spectrum and properties of the PTA. 
The SNR of the \CF\ is defined as
\be
\snrccf = \frac{\Chat_{00}}{\sigma_{\Chat}} = \frac{\psi\siggw^2}{\sigma_{\Chat}},
\ee
where the rms variation  $\sigma_{\Chat}$ is given in the Appendix
and includes the contributions from the GWs themselves along with 
uncorrelated red and white noise. 
For arbitrary combinations the SNR is
\be
  \snrccf &=& 
	\frac{\sqrt{\varrat\Np M}}{2}
	\!
	\left\{
	  w_{gg} + \zetaM w_{rr}  
	  + \frac
		{\left(w_{gg} + \zetaM^2 w_{rr} + 2\zetaM w_{gr}\right)} 
		{2\varrat(\Np-1)}
	  + \frac{\eta_M M}{\Nt} 
		\left[
			1 + \frac
				{(\eta_M/\Ns + 2 + 2\zetaM)} 
				{2\varrat(\Np-1)} 
		\right]
	\right\}^{-1/2}
	\!
	\!
	\!
	\!
	\!
	.
\label{eq:snr_all}
\ee
Simulations yield SNRs that agree with this expression to within statistical
errors. 
In Eq.~(\ref{eq:snr_all}) we have allowed for the division 
of the full time series into $M$ blocks and incoherent averaging
of the \CF\ for each block, reducing the variance by
$1/M$.   Both $\sigma_u$ and $\siggw$ depend implicitly  on
$M$,  as discussed below.  
The SNR depends on 
rms red and white noise levels through 
the variance  ratios
$\eta_M = \sigma_n^2 / \siggw^2$ and $\zetaM = \sigma_r^2 / \siggw^2$
that depend on the time span $T/M$
owing to the nonstationarity of the red noise and GW signals.
The number of blocks also enters in the leading coefficient in 
Eq.~(\ref{eq:snr_all}),
because the correlation estimate from each block is averaged and we assume that 
the estimates are statistically independent.  We have verified that
red processes with spectral indices  of four or less show this
statistical independence.   For steeper spectra, there is some
correlation between blocks.

For a dimensionless strain amplitude
spectrum $h_c(f) = A f^{\alpha_g}$, 
the spectrum of timing residuals 
$\propto f^{2\alpha_g-3}$ and the rms residual scales as 
$\siggw(T) \propto T^{x_g}$
with $x_g = 1-\alpha_G$ for $\alpha_g < 1$. For the GW background produced by 
merging SMBHs \cite[][]{2003ApJ...583..616J}, $\alpha_g=-2/3$ and $x_g = 5/3$.    
We will use a fiducial value $A = 10^{-15}~{\rm yr}^{-2/3}$.
Similarly, red timing noise has been characterized
with exponents $x_r \approx 2\pm 0.2$ corresponding 
to a spectrum $\propto f^{-5\pm0.4}$  
\cite[][]{2010ApJ...725.1607S}.
The dimensionless ratios then
become $\eta_M = \eta_1 M^{2x_g}$ and $\zetaM = \xi_1 M^{2(x_g-x_r)}$, 
where $\eta_1$ and $\xi_1$ are the values for the full-length time
series ($M=1$). 

The quantities $w_{gg}, w_{rr}$ and $w_{gr}$ are 
dimensionless correlation times that are defined in the Appendix.
For steep power-law spectra, they are of order
unity and independent of the data-span length owing to self-similarity.    
This same statement holds whether we consider the timing residual
model in Eq.~(\ref{eq:xmod}) to represent pre-fit residuals or those
after removing a polynomial to account for corrections
to  the spin parameters. 
Inclusion of astrometric sinusoidal terms in the fitting function with 
one year and half-year periods will induce some dependence of the
dimensionless scales on $T$ but with diminishing importance as $T\gg 1$~yr.  
The equivalent quantity for white noise is $1/\Nt$ because adjacent samples are
uncorrelated.

Smoothing (low-pass filtering) and decimation of the time series  by $\Ns$ samples before correlation 
reduces the rms of the white-noise.    We consider cases where 
$\Ns \ll N_t/M$ so that with blocking there is still a large number of 
samples per block.  We also consider the implied smoothing
time $\Ns T / N_t$ to be small enough that it does not
reduce the variance of the red processes over the block length $T/M$.


The simplest case, though unrealizable, is where only GWs contribute to
the timing residuals with the $e$ term perfectly correlated and the $p$ term
completely uncorrelated.    The SNR is 
\be
\snrccf =
        \frac{1}{2} 
	\left[
		\frac{\varrat\Np M/w_{gg}} { 1 + 1/2\varrat(\Np-1)} 
	\right]^{1/2}
	\approx \frac{1}{2} \sqrt{\frac{\varrat\Np M}{w_{gg}}},
\label{eq:snr_gws_only} 
\ee
the approximate equality holding when the number of pulsars
is very large, $\varrat\Np \gg 1$. 
The SNR grows as $\sqrt{\Np}$ and can become arbitrarily large with $\sqrt{M}$
subject to the requirement that the continuum approximation holds for 
arbitrarily small $T/M$.   
When white noise or red noise contribute, however,  
there is a distinct maximum in $S$ vs. $M$.

A number of other features are evident in Eq.~(\ref{eq:snr_all}), 
which has terms inside the curly brackets 
scaling as ${\cal O} (1), 1/\Np, 1/\Nt$ and $1/(\Np\Nt)$.
The number of TOAs, $\Nt$, is important only as long as the white noise
part of the residuals is sizable.  If less than other terms, the number
of TOAs --- and thus any cadence in acquiring them --- becomes unimportant.
A larger number of pulsars can reduce the effects of both the red and white
noise from non-GW contributions in addition to reducing the uncorrelated
``pulsar'' part of the GWs.  For very large $\Nt$ and $\Np$, the SNR reduces
to that for the GW-only case.

\begin{figure}
\begin{center}
\includegraphics[scale=0.40, angle=0] 
{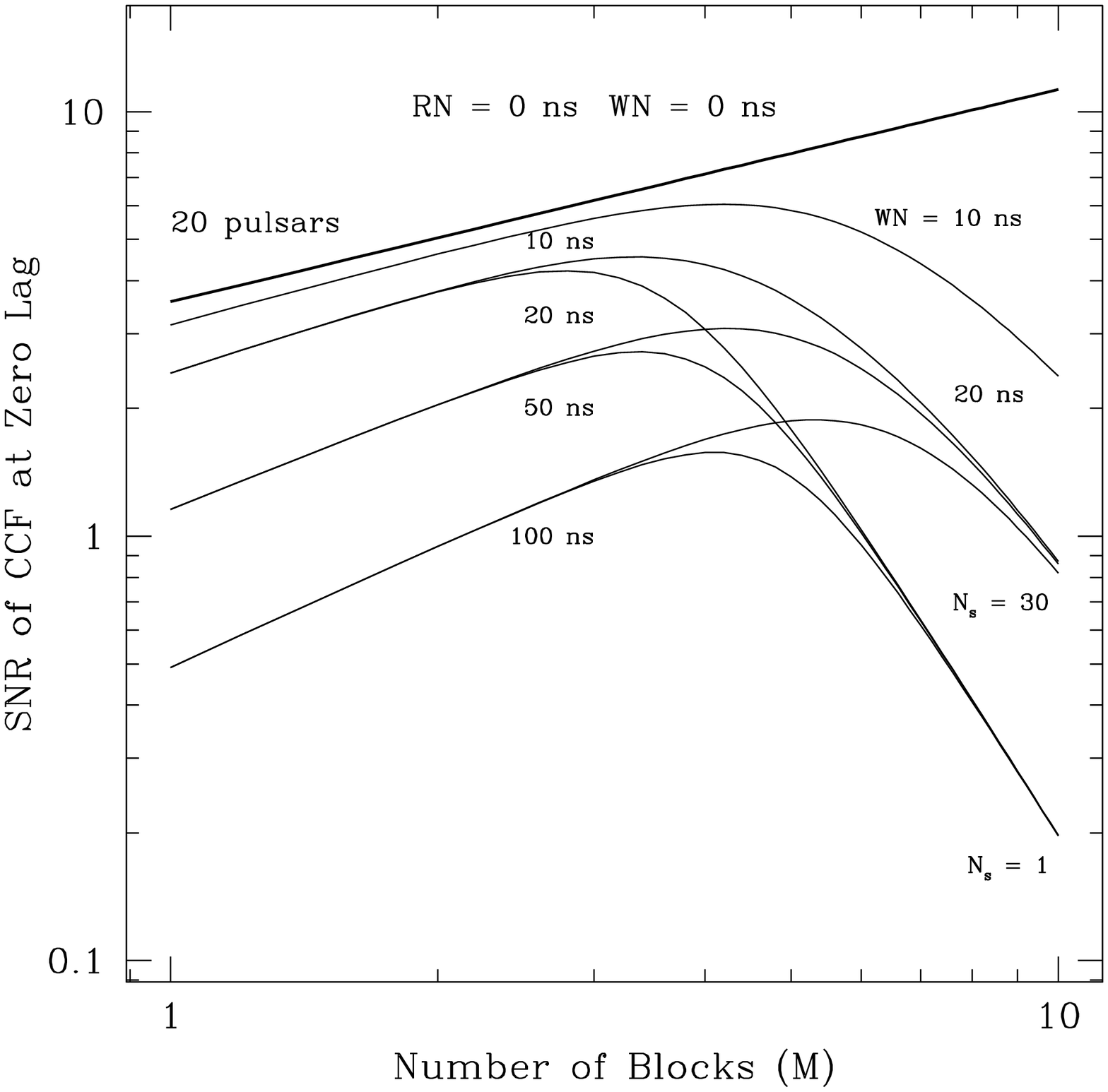}
\includegraphics[scale=0.40, angle=0] 
{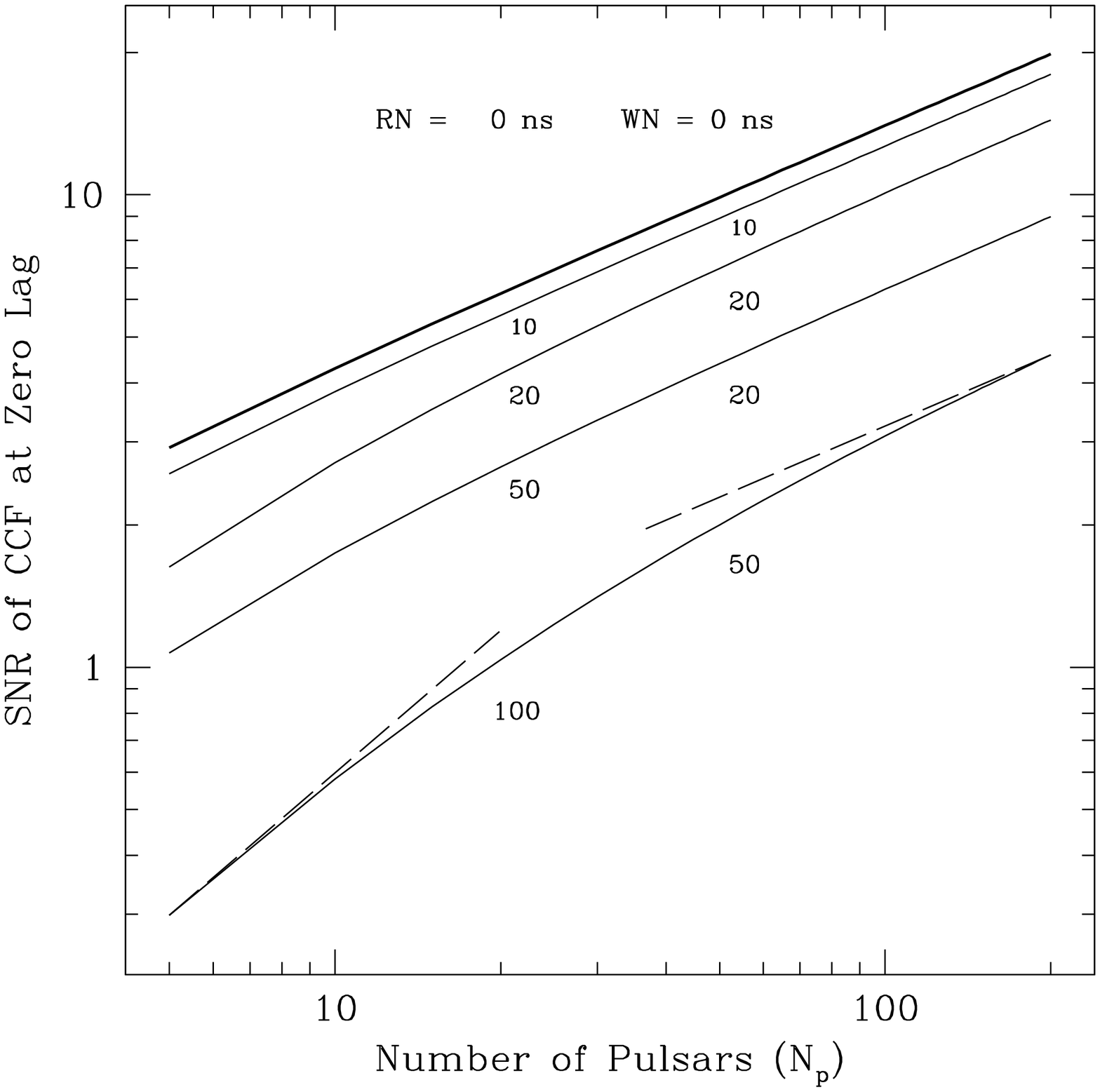}
\caption{
(Left)
SNR of the cross-correlation function 
{\em versus} number of blocks for cases where
the timing residuals include red noise from spin variations and/or ISM
perturbations along with GWs and white noise.   The plotted curves 
use $\sigma_r\propto T^{x_r}$ with $x_r = 2$ and are labeled with
the values for $T=5$~yr. 
For the six curves with 20~ns of white noise, three incorporate
smoothing of $\Ns = 30$ samples and the other three have no smoothing
($\Ns=1$).
\label{fig:SNR_PW_with_TN}
(Right)
SNR of the cross-correlation function {\em versus}  the number
of pulsars in the timing array sample.
Both red noise and white noise are included, as indicated.
Dashed lines indicate 
SNR~$\propto N_p$ and SNR~$\propto N_p^{1/2}$, which are the 
asymptotic scaling laws at low and high SNR, respectively.  
The plotted curves assume that $\sigma_r\propto T^{x_r}$ with $x_r = 2$  
and were calculated for $M=3$ blocks  and $\Nt = 10^3$ data points. 
\label{fig:SNR_vs_Np_TN_WN_pw}
}
\end{center}
\end{figure}

We illustrate these and other trends in Figure \ref{fig:SNR_PW_with_TN}.
In the left panel, 
the SNR is plotted against the number of blocks for
cases that include red and white noise added to the GW perturbation.
The plotted values are based on $\psi = 1$ and on a total of $N_t = 10^3$ TOAs over five years for 100 pulsars. 
We have dimensionalized the values for $\eta_M$ using 
$A = 10^{-15}$.
After a second order fit to a $T=5$~yr data span, the rms residual 
is $\siggw(T) \approx 20$~ns 
\cite[e.g.][]{2010ApJ...725.1607S}.

The curve for no noise (red or white) increases monotonically 
with $M$, but there is a distinct maximum SNR for non-zero noise that
separates the GW dominated and noise dominated regimes.  The optimal
number of blocks is $M\approx 3$ to 6 for the cases shown.  
In the noise-dominated regime, the SNR scales as
$\varrat\Np\sqrt{\Nt\Ns} M^{-2x_g}$, so 
smoothing improves the SNR but blocking does not.
When the SNR
is not  noise limited, it no longer depends on
$N_t$, so smoothing will have no effect.  This may be seen
in Figure~\ref{fig:SNR_PW_with_TN} (left panel) for the curves 
labeled 20, 50 and 100~ns
which converge at high SNR for both smoothing values shown, $\Ns = 1$ and 30.
In the GW dominated case,
there is no dependence on $\Nt$, $\Ns$,   
or on the scaling exponent,  $x_g$. 

Figure~\ref{fig:SNR_vs_Np_TN_WN_pw} (right panel) shows the SNR plotted against
the number of pulsars
used for several different values of white and timing noise.   The curves
in the figure were calculated for $M=3$  blocks and using a red noise
scaling $\sigma_r \propto T^{x_r}$ with $x_r = 2$. 
If red spin or ISM noise is absent ($\zetaM = 0$), 
$S\propto \sqrt{N_t}\Np$ for $S\ll 1$ is linear in the number of pulsars
(for large $\Np$) but then has 
a shallower dependence  $S\propto \sqrt{\Np}$ 
when $S$ is large enough to provide a confident detection. 
If we let $S_{\rm gw}$ be the SNR for the GW-only case in 
Eq.~(\ref{eq:snr_gws_only}), it can be shown that $S\ll S_{\rm gw}$ in
the white-noise limited case where the $\eta_M^2$ dominates other terms
in Eq.~(\ref{eq:snr_all}) and $S\propto \Np / \eta_M$.  Thus for $S$
large enough to correspond to a plausible detection, it is not likely 
to scale with $\Np$ as it does in the noise-limited 
regime but instead will scale as $S\propto \sqrt{\Np}$.   


\subsection{Comparison with Other Detection Approaches}
\label{sec:compare}

Our method uses the \CF\ as a test statistic and we calibrate it against
the SNR $S$ that can be related to theoretical GW spectra and sources
of noise.   
Our expression for $S$ in Eq.~(\ref{eq:snr_all}) is similar to 
Eq.~(12) of 
\citet[][]{2005ApJ...625L.123J},
who
define their detection significance as the SNR $S_J$ of a weighted correlation
quantity,  as discussed in \S~\ref{sec:detection}.   
However, their expression does not explicitly account for red and white
noise individually; instead a quantity $\chi$ is defined that measures
the degree of whiteness of the timing residuals.   Any non-white components
are assumed to arise solely from the GW background and not from any
additional spin-noise or ISM contribution.   As a consequence, we do not
expect the expressions for $S$ and $S_J$ to yield the same values
for the same PTAs. In addition,  it is assumed that $S_J$ has Gaussian 
statistics when there is no GW contribution, with a false-alarm fraction
calculated accordingly.   As we have shown, if red-noise contributes to
timing residuals,  the distribution of
$S$ is highly non-Gaussian both with or without a contribution from GWs.  

It is still instructive to compare nominal SNR values. 
For a pulsar timing array comprising $\Np = 20$~pulsars observed
$\Nt = 250$~times over $T=5$~years with an rms error
$\sigma_{n} = 100$~ns and no red noise, \citet[][] {2005ApJ...625L.123J}  
(their Figure 1)
find  
SNR~$\approx 2.8$ for a GW background of the same  form we have 
considered with $A = 10^{-15}$~yr$^{-2/3}$.
For the same PTA we find $S=2.9$ for $M=1$.  \citet[][] {2005ApJ...625L.123J}
obtain $S_J = 4.5$ using low-pass filtering (smoothing) and
pre-whitening.   Their low-pass filtering uses a high-frequency cutoff
$f_{hc} = 4/T$, corresponding to $\Ns = \Nt / 8$ in our notation.   
The blocking method we have analyzed shares some features similar to
pre-whitening.   Including smoothing and optimal blocking (which turns
out to be $M=1$ for this case), we obtain
$S = 3.1$.   As we have shown (Figure~\ref{fig:fdvsSNR}), this value is
sufficient to yield a high detection fraction ($\sim 0.99$) with a
corresponding fairly low false-alarm fraction (0.02).   Similar 
detection and false-alarm fractions are not available for the 
\citet[][] {2005ApJ...625L.123J} approach. We note also that our values
apply for the compact configuration of pulsars whereas those for
Jenet~et~al. are for an unspecified configuration. 

In a Bayesian treatment of GW detection, van Haasteren et al. (2009)  
cast detection in terms of parameter estimation and define the 
detection significance for the coefficient $A$ of the GW spectrum
using the SNR of $A$, $\mu/\sigma$, which we denote as $S_{\rm vH}$.
For a PTA with $N_p = 20$, $T=5$~yr and $\Nt =500$, their Figure~12
indicates $S_{\rm vH} \approx 2$~and~4 for  
$\sigma_{\rm WN}=100$~ns and 50~ns, respectively. 
For the same cases, we find $S=3.2$ and 3.8, respectively, with
no smoothing or blocking. \citet[][]{2010MNRAS.401.2372V} state that
their results are based on fixing all but one parameter ($A$) and thus
yield larger than expected SNRs.  We conclude that, nominally, our
SNRs are not inconsistent with those of \citep[][]{2010MNRAS.401.2372V}.
We emphasize, however, that the most meaningful comparison is of
detection and false-alarm fractions. 


\section{Climbing Mount Significance: Optimizing Detection}
\label{sec:mount}

Methods for increasing the detection signficance  
can be identified by using
Eq.~(\ref{eq:snr_all}) in various limiting cases. 
In the white-noise limited regime, the composite quantity
$Z_{\rm wn} = \Np\sqrt{\Nt\Ns}/\sigma_n^2$ 
can be inspected.    
When red noise dominates the SNR, the quantity 
$Z_{\rm rn} = \Np\sqrt{M}/\sigma_r^2$ 
is relevant.
In the GW-dominated regime, we have 
$Z_{\rm gw} = \sqrt{\Np M}$.  
\begin{enumerate}
\item
Increasing the number of pulsars $\Np$ helps in any regime,
though it has greater impact in the noise limited case (red or white).
Detection of GWs almost certainly will occur in the regime where the
SNR of the correlation-based detection statistic increases only
as the square-root of the number of pulsars.  
\item
In the white noise-limited regime, increasing 
$Z_{\rm wn}$
through a combination of more pulsars,
greater timing throughput, smoothing, and a decrease in timing error per
TOA will increase $S$.  
A detection fraction larger
than 0.9 combined with a false-alarm fraction $\lesssim 0.1$  
requires $S \gtrsim 2$.
\item 
Because
red noise is spectrally similar to the GW contribution and does not average
out significantly in the \CF\ because of its long correlation time, 
the primary means for increasing the
SNR is to sum over many pulsars and to use blocking. 
\item
In the GW-dominated regime, increasing the blocking $M$ and the number of 
pulsars $\Np$ are the only options.  As the figures show,  $M$ cannot be
increased arbitrarily because eventually the rms noise
in the interval $T/M$ will overwhelm the GW signal, which decreases
with smaller $T$.
\end{enumerate}

The sensitivity of a PTA to GWs of course improves with total observing span. 
In the white noise dominated case, these improvements are the greatest, 
with $S \propto T^{2x_g+1/2} \propto T^{23/6}$.   
However, a detection cannot be 
made in this regime where $S$ is small.    
In the regime where detections can 
be made, the longer observing span only enables a larger amount of 
sub-blocking so that $S \propto \sqrt{T}$.  

Of the factors we have discussed, the contributions
to red and white noise 
from interstellar refraction and diffraction 
\cite[][]{2010ApJ...717.1206C, 2010arXiv1010.3785C}
can be partly mitigated 
by using higher frequencies and by appropriate fitting across wide bandwidths. 
Larger telescopes and bandwidths can minimize radiometer noise but will 
have no effect on jitter.  Longer integration times are the only recourse 
for jitter but they also will minimize radiometer noise 

The most difficult hindrance to overcome is red noise from spin variations
and from any residual ISM effects that cannot be corrected.  
The range of red timing noise levels
in MSPs is not known definitively but our recent assessment 
\cite[][]{2010ApJ...725.1607S}
suggests that it is larger
in objects with larger spin-frequency derivatives.   
Latent red noise may emerge in many MSPs 
when more sensitive and longer time-span observations are obtained.    
If so, greater timing throughput will be needed to time more MSPs as well
as to increase the observing time per pulsar to reduce white noise
sufficiently for detection.

\subsection{Implications}
\label{sec:imps}

The GW dominated regime provides the largest SNR  and thus
indicates the absolute minimum number of pulsars needed to make a detection.
Defining a 
threshold $S > S_{\rm min}$, the number of pulsars
required is 
\be
\Np \ge 4{\rm S_{min}^2} w_{gg} / \varrat M.
\ee 
For $\varrat = 1$, $M=1$, and $w_{gg} = 0.4$ a minimum SNR of 2 requires 
$\Np=6$ pulsars,
as seen in Figure~\ref{fig:SNR_PW_with_TN}.  With no timing noise or 
white noise, $M$ can be made arbitrarily large (subject to sampling rates). 
However,  even optimistic values
of white noise and red noise (e.g. 20 and 20~ns, respectively) 
indicate that $S$ is maximized for
$M\approx 2$~to~3, so $\Np \gtrsim 7/\psi$ is needed for $M=3$.   The actual variance
of the ``Earth'' part of the GW background could have $\psi$ much larger or
smaller than unity, so detection of GWs with a small number of pulsars may be marginal.
A larger threshold ${\rm SNR_{min}} = 5$ will require 
$\Np\gtrsim 25/\varrat$ for $M=3$. 
Our results suggest that an increase in the cadence of timing measurements 
to  increase the total number of TOAs for each object ($\Nt$) may be 
a necessary but insufficient course to take for GW detection.

To obtain our results, we have assumed that the Earth term $e(t)$ 
is the same for all pulsars in a hypothetical spatial 
configuration. 
For realistic distributions of pulsars on the sky, 
the correlation amplitude will be reduced by 
$\overline{\zeta^2} \approx 0.6$, thus increasing the number of required 
pulsars by about a factor of $1/(\overline{\zeta^2})^2 \approx 3$. 
For PTAs with 
a range of red-and-white-noise levels,  
weights for each pulsar can be introduced
in the double sum in Eq.~(\ref{eq:Chat}).   For a nominal level of
white noise, for example, with some objects having smaller and others
larger rms values,  a weighted sum will yield a larger SNR than the
case where all objects have the same rms white noise.  If we take the
nominal value as the {\em minimum} in the sample, however, the optimally
weighted $\Chat$ will have lower SNR.

Currently only two MSPs (J1713+0747, J1909$-$3744)
are known to have rms timing residuals
less than 50~ns over a 5-yr interval 
\cite[][]{2009astro2010S..64D}
and one other less than 100~ns, J0437$-$4715 
\cite[][]{2010HiA....15..233M}.
An aggressive campaign 
is needed to find more MSPs with timing noise substantially less
than 100~ns in a 5-year span.  
The timing noise scaling law of 
\citet[][]{2010ApJ...725.1607S}
suggests that such MSPs will have small spindown rates   and they may be
less luminous if the radio beam luminosity correlates
with the energy loss rate. This implies that MSP surveys
may need to be more sensitive than at present.

The best strategy is to identify $\sim 20$ 
``super''-stable MSPs with rms timing noise of 20~ns or less over time
spans of  5 years or more if a detection threshold $S_{\rm min}=2$ is considered
sufficient for detection.    
A larger $S_{\rm min} = 5$ requires $\sim 50$ objects. 
However, if no such super-stable objects exist and MSPs more typically
have 20~ns rms timing noise or larger over 5 years,  
many more MSPs will need to be timed, perhaps exceeding 100 MSPs.

Even if the super-stable regime applies, once a detection is made,
possibly using existing telescopes to time 
known stable objects along with any new discoveries in the near term,
a more detailed analysis of the GW spectrum will be desired and that certainly
will require a much larger set of MSPs and overall greater throughput of
the timing. 



Each MSP needs a careful error budget analysis.  This would include
a detailed characterization of the
red and white noise levels, including a breakdown of each from different
physical causes.  
The two kinds of noise can 
be distinguished through appropriate use of the structure function of 
timing residuals 
\cite[e.g.][]{1985ApJS...59..343C}. 
Departures from white noise need to be characterized according to amplitude
and spectrum and classified as  contributions from
red noise of any kind, from changes in instrumentation, which can cause jumps in pulse phase
between epochs. 
A change-point analysis \citep[e.g.][Chapter 5]{NumBayes}  on timing residuals can identify the amplitudes of 
such jumps whether or not their occurrence epochs are known. 
MSPs with significant red noise that is demonstrated to be from non-GW
causes should be rejected because they do not contribute to the
sensitivity of a PTA to a stochastic background of GWs. 

\section{Discussion}
\label{sec:discussion}

The main results of our paper are as follows.

The cross-correlation function is the primary statistic that we consider for a hypothetical 
pulsar distribution that yields the maximum possible signal to noise ratio, all else being equal. 
For a realistic distribution, the equivalent quantity would be a weighted sum similar to
the quantity $\rho$ defined by 
\cite[][]{2005ApJ...625L.123J}, but with a time-lag argument included. 
The \CF\ has amplitudes over an ensemble that have a positive skewed distribution that influences
the detection and false-alarm fractions.    We have taken this into account in our analysis and we
also suggest that sub-dividing the entire span of timing residuals into $M\approx 3$ sub-spans
for each pulsar will reduce the skewness and increase the statistical significance.
Such blocking is equivalent to high-pass filtering the data.
It requires a large-enough cadence for TOA measurements that there is an ample 
number of samples within
each interval of length $T/M$. 

The number of pulsars needed for a likely
detection of nanohertz GWs depends strongly on the levels of white noise and 
especially the red noise in the data.   In related papers 
\cite[][Shannon \& Cordes, in preparation]{2010arXiv1010.3785C}
we have shown that
both kinds of noise are likely to be present due to torque noise in
the pulsar, magnetospheric motions of emission regions, and interstellar
plasma phenomena.   

Red and white noise timing residuals dramatically alter the
achievable signal to noise ratio of the \CF.
When residuals are white noise dominated,
improvements can be made by 
increasing the net integration time per pulsar or by smoothing individual measurements
over a time shorter than the smallest GW period that is likely to be 
identified.  Marginal gains can also be made from blocking of the data. 

If, however, the detection statistic
is dominated by red noise with a power spectrum similar to that
of the GW power spectrum, smoothing or other increases in net integration time per
TOA will not help.  
The best recourse is to increase the
number of pulsars in the pulsar timing array.     

We have shown that the correlated GW signal contributes variance
that can differ markedly from the ensemble value if the GW signal
has a steep power-law spectrum, like that expected from 
merging of supermassive black holes.   This stochasticity of the
sample variance can either greatly enhance or diminish the chances
of detecting the signal.  The skewness of the correlation function's
signal to noise ratio is an important factor in assessing detection
and false-alarm statistics.   The skewness is reduced when time 
series are divided into sub-blocks that are analyzed separately and then
combined.

As mentioned in the previous section, 
an important action is the characterization of the timing error
budget for each MSP.   

We consider it likely that 50~to~100 spin-stable MSPs are needed
to fully characterize GWs at nanohertz frequencies, including
a secure detection followed by detailed characterization.  
A minimum of 20 MSPs is needed
for a plausible detection under optimistic red and white noise levels, 
as described in this paper.  
A sample that is distributed on the sky will increase
this number by $\sim 60$\%   and verification with a 
completely independent sample will require
another doubling. 
The program going
forward therefore requires an aggressive search campaign to discover more
MSPs and to identify the most spin-stable objects.  It is possible that
the most stable objects are also those with smaller radio luminosities.
The scaling law for red noise identified by 
\citep[][]{2010ApJ...725.1607S}
implies
that objects with larger spin-down rates have larger noise levels.   While not known
for certain, the radio luminosity likely also is larger for these objects. 
In addition, further study of spin noise in MSPs is needed to ascertain whether it can
be mitigated, as suggested by 
\citet[][]{2010Sci...329..408L}.

We thank members of the NANOGrav collaboration for useful comments on 
this work.
Our work was supported by the NSF through a subaward to Cornell University
from  
Partnerships for International Research and Education award 0968296 
to West Virginia University. 

\newpage
\appendix
\section{Appendix}
\label{sec:app}

\subsection{Definitions and Correlation Time Scales}

The GW perturbations induced in a pair of pulsars
are correlated according to their angular separation
\citep{1983ApJ...265L..39H}.
Following 
\citet[][]{2005ApJ...625L.123J}
and others, 
we define an estimator for the  angular and temporal correlation function 
as an integral over time and a sum over the
$\NX = \Np(\Np-1)/2$ unique pairs of pulsars,
\be
\Chat(\theta, \tau) =  
	\frac{1}{\NX(\theta)}
	\sum_{i,j: \theta} 
	\frac{1}{T}
	\int_0^Tdt\, x_i(t) x_j(t+\tau),
\ee
where pairs are summed such that the separation angle
$\theta_{ij}$ between the $i^{th}$ and $j^{th}$ pulsars is within an angular
bin centered on $\theta$.  In practice, the time integral is a sum over
discrete time series, but it is more useful for our analysis to use
continuous notation.   It is easy to collapse our continuous-time result
to the discrete-time case as discussed below.

In the main text we discuss the zero-lag correlation
$\Chat_{00} = \Chat(0, 0)$ and its signal-to-noise ratio,
$S = \Chat(0,0)/\sigma_{\Chat}$. 
For a general $u(t)$, the rms $\Chat_{00}$ is
\be
\sigma_{\Chat} = \frac{2}{\sqrt{\Np}}
	\left[
	\varrat\siggw^2 \sigma_u^2 w_{eu} 
		+ \frac{\sigma_u^4 w_{uu}}{2(\Np-1)}
	\right]^{1/2},
\label{eq:sigCu}
\ee
where $\psi \siggw^2$ is the mean square of $e(t)$ over the time interval and
$\psi\approx 1$ takes
into account that $e(t)$ is a single realization of the GW process while we define $\siggw^2$ to be the
ensemble-average variance (see further discussion below). 
The uncorrelated noise $u$ has variance $\sigma_u^2$; $w_{eu}$ and $w_{uu}$ are dimensionless correlation
times of order unity that are discussed below. 
If there are no GWs, we substitute
$\sigma_u^4 w_{uu} \to \sigma_r^4 w_{rr} + (\sigma_n^4 + 2\sigma_r^2\sigma_n^2)/\Nt$,
where $\Nt$ is the number of discrete samples in $[0,T]$. 
Using additional dimensionless correlation times 
for the no-GW case, we obtain
\be
\sigma_{\Chat} = \frac{\sqrt{2}}{\Np(1-1/\Np)^{1/2}}
	\left(\sigrn^4 w_{rr} + \frac{\sigwn^4 + 2\sigrn^2\sigwn^2}{\Nt} \right)^{1/2}.
\label{eq:sigC_noGWs}
\ee
The rms correlation scales as the inverse of the number of pulsars and 
is independent of the number of time samples when the red noise dominates.
When GWs are significant, the dependence on $\Np$ is slower, for large $\Np$ scaling as 
$\sigma_{\Chat} \approx  2\siggw^2 \sqrt{w_{ep} / \Np}$. 
We have verified these scaling laws using
simulations with different combinations of white and red noise and number
of pulsars.  

For arbitrary combinations of GWs, red, and white noise, we expand
$u = p + r + n$ and use corresponding variances and correlation functions 
to get
\be
  \sigma_{\Chat} &=& 
	\frac{2\siggw^2\sqrt{\displaystyle\varrat}}{\sqrt{\Np M}}
	\!
	\left\{
	  w_{ep} + \zetaM w_{er}  
	  + \frac
		{\left(w_{pp} + \zetaM^2 w_{rr} + 2\zetaM w_{pr}\right)} 
		{2\varrat(\Np-1)}
	  + \frac{\eta_M M}{\Nt} 
		\left[
			1 + \frac
				{(\eta_M/\Ns + 2 + 2\zetaM)} 
				{2\varrat(\Np-1)} 
		\right]
	\right\}^{1/2}
	\!
	\!
	\!
	\!
	\!
	.
\label{eq:sigC_all}
\ee

The dimensionless
time scales used in the expressions for $\sigma_{\Chat}$
result from  double integrals that comprise the variance of the correlation function,
\be
w_{ab} =  
        T^{-2} \iint\limits_0\limits^{~~~~ T} dt\,d\tp\, 
	\rho_{a}(t,\tp) \rho_{b}(t,\tp),
\label{eq:wfactor}
\ee
where the factors in the integrand are normalized correlation functions
for individual terms defined in Eq.~(\ref{eq:xmod2}), 
$\rho_{a}(t,\tp)$, with
$a = e, p, r, n$.   
The correlation functions have two arguments  
because most of the
processes we consider have nonstationary-like statistics over finite
time intervals.   
We have $\sigma_{a}^2(T) = T^{-1}\int_0^T dt\, \langle a^2(t) \rangle$
and 
$\rho_{a}(t,\tp)\equiv\langle a(t) a(\tp)\rangle/\sigma_x^2(T)$.  

We account for the fact that $e(t)$ represents a single
realization of the GW signal.
Under the assumption that all pulsars in the sample are
in the same direction,  
$e(t)$ is identical for all lines of sight.
By contrast, the GW signal at each pulsar's location, $p(t)$,
is different for each LOS so that a set of $\Np\gg 1$ pulsars
samples a range of variances that are well represented by the 
ensemble average.  Therefore we write
\be
	\sigma_e^2(T) \rho_e(t,\tp) 
		\equiv \varrat \sigma_p^2(T) \rho_p(t,\tp)
		\equiv \varrat \siggw^2(T) \rho_g(t, \tp),
\ee
where $\varrat$ is the ratio of the realization variance to the ensemble
variance for the GWs. 

Even though $e(t)$ represents a single realization  
while the correlation estimator
uses $\Np\gg 1$ independent realizations of $p(t)$, we approximate both as 
having the same correlation function and same dimensionless correlation time,
$w_{gg}$.  We therefore let $w_{ep} \approx w_{pp} \approx w_{gg}$
and $w_{er} \approx w_{pr} \approx w_{gr}$.

The white-noise correlation 
$\rho_n$ decorrelates on a time scale of one sample in discretely sampled data.
Therefore the factors 
$w_{en}, w_{pn}, w_{rn}$, and $w_{nn}$, 
all become $ 1/\Nt$, the reciprocal of the number of time samples. 
This approach is a good representation of timing residuals
with irregular, discrete sampling where $N_t$ is large
enough to sample adequately the red processes, including the GW signal. 

In contast to white noise, 
the steep red power spectra for $e, p$ and $r$ 
yield dimensionless time scales $\gg 1/\Nt$.
Using simulations like those described below
 we find $w_{ep} \approx 0.36$ 
for red noise created with an $f^{-13/3}$ spectrum   
after removal of a second order polynomial. 
For red noise consistent with timing noise in pulsars ($\propto f^{-5}$),
we obtain $w_{er} \approx 0.44$.   For a flat spectrum (white noise),
we verify that $w_{nn} = 1/\Nt$ to within statistical errors in simulations.

Eq.~(\ref{eq:sigC_all}) incorporates smoothing of the original time series by $\Ns$ samples
and for blocking of the total data span of $\Nt$ samples into subintervals of
$\Nt/M$ samples corresponding to a time interval $T/M$.  Each subinterval is processed separately
and the correlation estimates are summed, reducing the rms $\sigma_{\Chat}$ by a factor $1/\sqrt{M}$. 
Clearly, any smoothing and blocking must yield a net number of samples per subinterval  to be large
enough to allow for a polynomial fit.   We have also used dimensionless variance ratios,
$\eta_M = \sigwn^2 / \siggw^2(T/M)$ and $\xi_M = \sigrn^2(T/M) / \siggw^2(T/M)$. We have explicitly 
indicated that the variances for the GWs and red noise are functions of the data span length, $T/M$.  
We discuss these in detail in the main text.

Figure~\ref{fig:psihists} shows 
histograms of the rms timing perturbations
before and after fitting a quadratic function 
for processes with spectral indices $y = 2, 4, 6$ (which correspond roughly to
random walks in spin phase, frequency and frequency derivative)
and $y = 13/3$, the value expected from a GW background produced by
merging supermassive black holes (SMBHs).  
Histograms are also shown for pure white noise ($y=0$). 
The ratio of rms values  equals $\sqrt{\varrat}$ for the pre-fit case and 
is proportional to $\sqrt{\varrat}$ for the post fit case. 
The ratio varies by more
than an order of magnitude before fitting 
 but covers a somewhat smaller range after fitting.  

\setcounter{figure}{0}
\begin{figure}[h!]
\begin{center}
\includegraphics[scale=0.45, angle=0]
{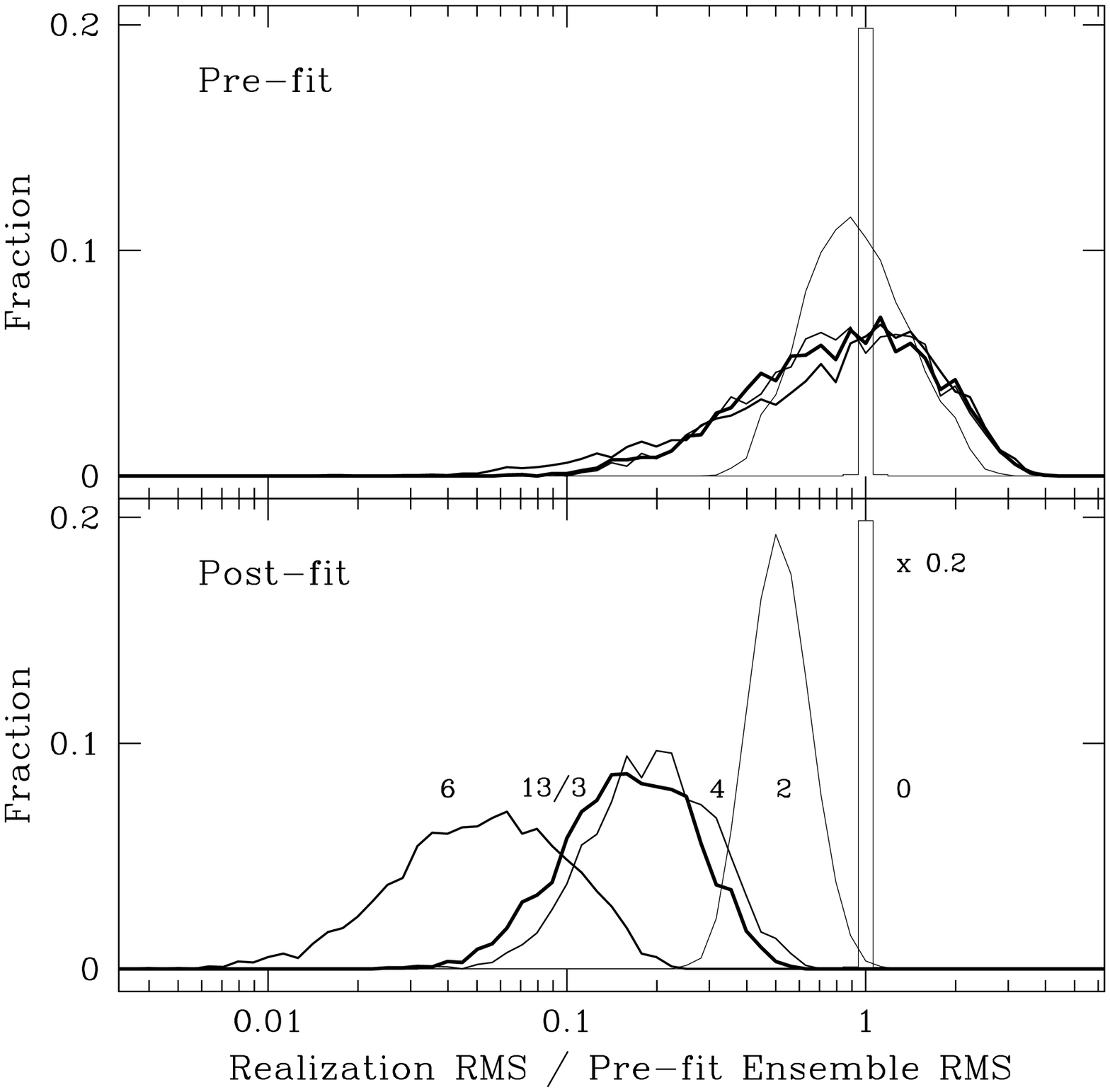}
\caption{
Histograms of the rms TOA for red processes with power-law spectra
$\propto f^{-y}$ with $y = 0, 2,4,6$ and $13/3$, as labelled in the
bottom panel. 
The top and bottom
panels show results before and after removing a second-order polynomial.
The rms values are normalized by the ensemble rms of the pre-fit
time series.
\label{fig:psihists}
}
\end{center}
\end{figure}


\subsection{Generation of Simulated Time Series}

We generate realizations of red noise by shaping complex white noise in the frequency domain and
performing an inverse discrete Fourier transform.    We fill an array that includes Fourier components
with periods that are four times longer than our desired time series so that low-frequency components
are not underestimated.  We then select 1/4 of the time series.   To suitably mimic the analysis of
pulsar timing data, we subtract a straight line whose end points equal the first and last data points.
This accounts for the fact that prior to doing a least-square fit to timing data, a preliminary timing model
is first removed.  In this way we compare pre-and-post fit variances that are close to representing
those that would result in actual applications. 

\subsection{Non-Gaussianity of the Zero-lag \CF\ }

In the main text we describe the skewness of the distribution of $\Chat_{00}$ toward positive values.
The skewness is generic for time series that include a red-noise component.  Two effects lead to this result.
First, inspection of Eq~(\ref{eq:Chat}) shows that the correlation function for a single pair of pulsars
is an integral (or sum) of the products of two time series.   The CLT will apply if each time series
includes many independent fluctuations over the interval $[0,T]$.   The sum over all pairs will also
satisfy the conditions for the CLT and $\Chat_{00}$ will have a Gaussian distribution.  For red noise
processes, however, each time series is dominated by of order only one fluctuation, so the CLT will not apply
to the single-pair integral.   The second effect is that when the CLT does not
apply to the CCF for a single pair it also does not apply to the sum over 
all pairs, in part
because a given time series contributes to $\Np -1$ terms in the sum and 
the terms are not independent.

\end{document}